\documentclass[lettersize,journal]{IEEEtran}
\usepackage{amsmath,amsfonts}
\IEEEoverridecommandlockouts
\usepackage{cite}
\usepackage[caption=false,font=normalsize,labelfont=sf,textfont=sf]{subfig}
\usepackage{float}
\usepackage{stfloats}
\usepackage{fancyhdr}
\usepackage{color}
\usepackage{amsmath}
\usepackage{graphicx}
\graphicspath{{./figures/}}
\usepackage{algorithm}
\usepackage{xcolor}
\usepackage{algorithmic}
\usepackage{epsfig}
\usepackage{txfonts}
\usepackage{epstopdf} 

\begin{document}
	\title{Green Beamforming Design for Integrated Sensing and Communication Systems: A Practical Approach Using Beam-Matching Error Metrics	
	}
	
	\author{\IEEEauthorblockN{
		      Luping Xiang, \emph{Member, IEEE},
			Ke~Xu, \emph{Student~Member, IEEE},
			Jie~Hu, \emph{Senior~Member, IEEE},
			and~Kun~Yang}, \emph{Fellow, IEEE}
			\vspace{-0.5 cm}\\

        \thanks{Luping Xiang, Ke Xu, and Jie Hu are with the School of Information
        and Communication Engineering, University of Electronic Science and
        Technology of China, Chengdu 611731, China, email: luping.xiang@uestc.edu.cn, 202121010634@std.uestc.edu.cn, hujie@uestc.edu.cn. \textit{(Corresponding author: Jie Hu.)}}
        \thanks{Kun Yang is also with the School of Computer Science and Electronic Engineering,
        University of Essex, Essex CO4 3SQ, U.K., e-mail: kunyang@essex.ac.uk}
	}
	
	\maketitle
	
	\thispagestyle{fancy} 
	\lhead{} 
	\chead{} 
	\rhead{} 
	\lfoot{} 
	\cfoot{} 
	\rfoot{\thepage} 
	\renewcommand{\headrulewidth}{0pt} 
	\renewcommand{\footrulewidth}{0pt} 
	\pagestyle{fancy}

    \rfoot{\thepage} 

	\begin{abstract}
In this paper, we propose a green beamforming design for the integrated sensing and communication (ISAC) system, using beam-matching error to assess radar performance. The beam-matching error metric, which considers the mean square error between the desired and designed beam patterns, provides a more practical evaluation approach. To tackle the non-convex challenge inherent in beamforming design, we apply semidefinite relaxation (SDR) to address the rank-one relaxation issue, followed by the iterative rank minimization algorithm (IRM) for rank-one recovery. The simulation results showcase the effectiveness of our proposed optimal beamforming design, emphasizing the exceptional performance of the radar component in sensing tasks.
	\end{abstract}
	\section{Introduction}
In recent years, the fusion of sensing and communication, known as Integrated Sensing and Communication (ISAC), has emerged as a crucial element for sixth-generation wireless networks, addressing the issue of spectrum conflicts between communication and radar systems \cite{ISACsurvey}. The ISAC system employs a jointly designed waveform to convey information to communication users while ensuring that sensing services are provided for radar targets. ISAC plays a significant role in Vehicle-to-Everything (V2X), Internet of Everything (IoE), and smart city applications.

Optimal ISAC design approaches can be classified into three categories: communication-centric design, radar-centric design, and integrated design. The communication-centric design utilizes existing communication waveforms to carry out both communication and radar tasks, such as Orthogonal Frequency Division Multiplexing (OFDM) \cite{OFDM-center} and Orthogonal Time Frequency Space (OTFS) \cite{OTFS-center}. In the radar-centric design, information is embedded into traditional radar waveforms, for instance, Linear Frequency Modulation (LFM) \cite{LFM}. Currently, integrated waveform design aims to design a brand new dual-functional waveform. 
Comparing the radar-centric and the communication-centric, the integrated design break through the limitations of the original waveform, which can  offer increased Degrees of Freedom (DoFs) and adaptability to get better performance of the ISAC system \cite{integrated-design}.

There have been a number of works on the integrated design, but different optimisation result can be obtained based on different evaluation metrics.
Presently, different evaluation metrics for radar systems encompass beam gain, beam-matching error, Cramér-Rao Bound (CRB), and mutual information (MI). The initial metric, beam gain, is widely utilized to represent the signal strength at a specific angle of interest. For example, in \cite{UAV_beamGain}, the authors jointly design a beamforming and an unmanned aerial vehicle (UAV) trajectory to meet beam gain constraints and other communication constraints at all times. The beam-matching error refers to the mean square error (MSE) between a predetermined desired beam pattern and the designed beam pattern. In \cite{Beam_match}, the optimal beamforming is explored to serve multiple radar targets and communication users to minimize beam-matching error. CRB serves as the lower bound for the variance of unbiased parameter estimation. CRB for radar targets and signal-to-interference-plus-noise ratio (SINR) for communication users are also applied in ISAC beamforming design \cite{CRB}. Radar MI describes the amount of target information that can be transmitted to the receiver. MI between the radar channel, which includes the target response, and the received signal has been employed to assess the OFDM beamforming design in the ISAC system \cite{MI}.

However, the aforementioned ISAC beamforming design has mainly focused on improving the tradeoff between sensing and communication performance, without taking into account power consumption and the impact of the beam width.  In contrast, the study in \cite{EE_CRB} investigated an energy-efficient beamforming design for ISAC, maximizing the ratio of communication rate to total power while satisfying radar CRB constraints. Another study in \cite{Power_Min_MI} focused on power minimization-based OFDM radar beamforming design for ISAC, using MI to evaluate radar performance. However, to the best of our knowledge, optimal ISAC beamforming design that considers power minimization under the beam-matching error metric has not been investigated yet.

Motivated by the need to address existing research gaps, we propose a green ISAC beamforming design that considers power minimization. Different from other evaluation metrics, the beam-matching error can accurately represent the estimation capability, which is more practical to assess radar performance. Moreover, the distinctions between beam-matching and beam gain metrics in the green ISAC beamforming design will be elucidated. The imapct of the beam width will be also discussed as a focus in this paper. The main contributions of this work can be summarized as follows:

	\begin{itemize}
		\item We propose a green beamforming design for ISAC systems, where the base station (BS) designs beamforming matrix to achieve the desired information rate of the communication users while satisfying radar sensing constraints through the use of a beam-matching error matrix.
		\item To address the non-convex problem of beamforming design, we employ semidefinite  relaxation (SDR) to tackle the rank-one relaxation issue. Subsequently, the iterative rank minimization algorithm (IRM) is utilized for rank-one recovery.
		\item Simulation results showcase the effectiveness of our proposed green beamforming design, emphasizing the exceptional performance of the radar component in sensing tasks.
	\end{itemize}

    The remainder of this paper is structured as follows: Section II presents the system model, followed by the green beamforming design in Section III. Section IV provides simulation results, and the paper concludes in Section V.
\begin{figure}[!t]
		\centering
		\includegraphics[width=0.35\textwidth]{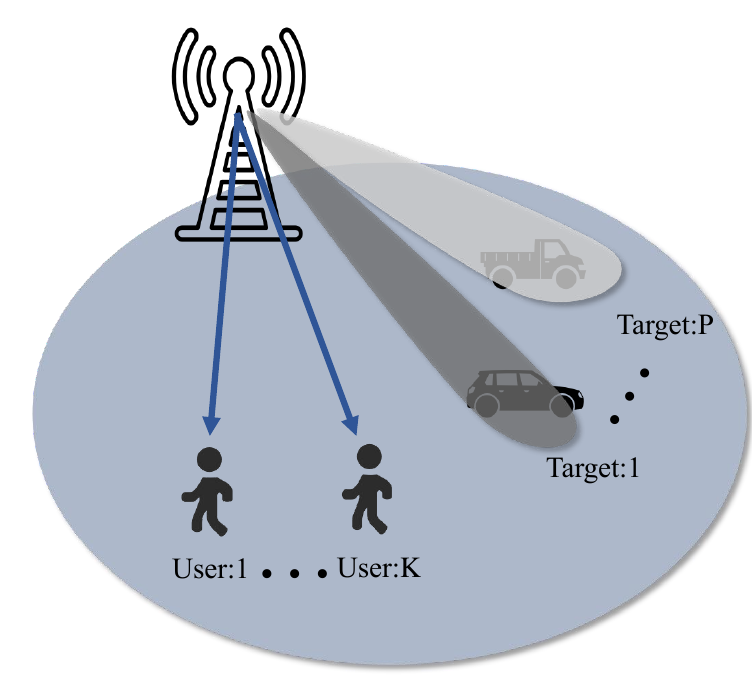}
		\caption{ISAC downlink system model.} \label{fig:system.pdf}
	\end{figure}
\section{System Model}

\label{sec.transmitter}
We consider the ISAC downlink system illustrated in Fig. 1, in which the base station (BS) features a uniform linear array (ULA) with $N$ antennas, catering to $K$ communication users and $P$ radar targets, where $s_{k}$ denotes the information signal for the $k$-th user, with $k=1,2,\dots, K$. The associated transmit beamforming vector is $\textbf{\emph{w}}_k \in \mathbb{C}^{N \times 1}$. The information signal for the $k$-th user $s_k$ is presumed to be Gaussian random variables with zero mean and unit variance, fulfilling $\mathbb{E}(s_k s^H_k)=1$. $\mathbf{s}_\text{0}$ represents the dedicated radar sensing signal, generated using pseudo-random coding \cite{randomcoding} with zero mean and covariance matrix $\textbf{\emph{R}}_d=\mathbf{E}(\mathbf{s}_\text{0}\mathbf{s}^{H}_\text{0})\succeq 0$. By combining the communication and radar signals, the transmitted signal $\textbf{\emph{x}}\in \mathbb{C}^{N\times 1}$ can be expressed as:

\begin{align}
		\begin{split}
		\textbf{\emph{x}}=\sum_{k=1}^{K}{{\textbf{\emph{w}}}_k s_k}+\textbf{\emph{s}}_{\text{0}}.
		\end{split}
\end{align}
Thus, the average transmit power can be denoted as:
\begin{align}
		\begin{split}
		\mathbb{E}(||\textbf{\emph{x}}||^2)=\sum_{k=1}^{K}{||{\textbf{\emph{w}}}_k||^2}+\text{tr}(\textbf{\emph{R}}_d),
		\end{split}
\end{align}
where $||\ast||^2$ denotes the Euclidean norm of a complex
vector.
In the ISAC beamforming design, we consider the line-of-sight (LOS) channel model as in \cite{UAV_beamGain}. The channel vector from the BS to the $k$-th user can be expressed as:
\begin{align}
		\begin{split}
		{\textbf{\emph{h}}}_k(\theta_k)=\beta_k(d_k){\textbf{\emph{a}}}(\theta_k),
		\end{split}
\end{align}
where $\beta_k(d_k)=\frac{G_{\text{T}}G_{\text{R}}{\lambda}^2}{(4\pi)^2 d_k^2}$ denotes the large-scale fading, with $G_{\text{T}}$ and $G_{\text{R}}$ representing the transmit and receive gain, respectively, and $\lambda$ denotes the electromagnetic wavelength. The steering vector ${\textbf{\emph{a}}}(\theta_k)$ is defined as:
\begin{align}
		\begin{split}
		{\textbf{\emph{a}}}(\theta_k)=\left[1,e^{j2\pi \frac{d}{\lambda}\cos{\theta_k}},\dots,e^{j2\pi \frac{d}{\lambda}(N-1)\cos{\theta_k}}\right]^{\text{T}},
		\end{split}
\end{align}
where $d=\frac{1}{2} \lambda$ denotes the spacing  between two adjacent antennas.
The received signal at the $k$-th user is given by:

\begin{align}
		\begin{split}
		\textbf{\emph{y}}_k={\textbf{\emph{h}}}_k^{\text{H}}(\theta_k) \textbf{\emph{x}}+n_k={\textbf{\emph{h}}}_k^{\text{H}}(\theta_k)(\sum_{k=1}^{K}{{\textbf{\emph{w}}}_k s_k}+\textbf{\emph{s}}_{\text{0}})+n_k, \label{received signal}
		\end{split}
\end{align}
where $n_k\sim \mathcal{C N}(0, \sigma^2_k)$ represents the additive white Gaussian noise (AWGN) at the receiver of the $k$-th user. From the Eq. \eqref{received signal}, it can be observed that the $k$-th user receives the desired signal $s_k$, interference from  other users  $\{s_i\}_{i\neq k}$, radar signal $s_\text{0}$ and noise $n_k$. Therefore, the corresponding SINR of $k$-th user can be formulated as:

\begin{align}
		\begin{split}
		\gamma_k=\frac{|{\textbf{\emph{h}}}^{\text{H}}_k(\theta_k){\textbf{\emph{w}}}_k|^2}{\sum_{i=1,i\neq k}^{K}{|{\textbf{\emph{h}}}^{\text{H}}_k(\theta_k){\textbf{\emph{w}}}_i|^2}+{\textbf{\emph{h}}}^{\text{H}}_k(\theta_k)\textbf{\emph{R}}_d{\textbf{\emph{h}}}_k(\theta_k)+\sigma_k^2}.
		\end{split}
\end{align}
Hence, the information rate of $k$-th user can be characterized by the classic Shannon-Hartly capacity, expressed as follows:
\begin{align}
		\begin{split}
		R_{k}=\log_2\left({1+\gamma_k}\right).
		\end{split}
\end{align}

We assume that the angle of the $p$-th radar target is denoted by $\theta_p$, with $p=1,\dots,P$. Then the corresponding radar channel can be denoted as ${\textbf{\emph{h}}}_p(\theta_p)$.
In terms of radar sensing, information signals are regarded as supplementary radar gain. Consequently, the beam gain can be expressed as:
\begin{align}
       \begin{split}
		 \xi^{\text{\uppercase\expandafter{\romannumeral1}}}(\theta_p)={\textbf{\emph{h}}}^{\text{H}}_p(\theta_p)(\sum_{k=1}^{K}{{\textbf{\emph{w}}}_k {\textbf{\emph{w}}}^{\text{H}}_k}+\textbf{\emph{R}}_d){\textbf{\emph{h}}}_p(\theta_p).
        \end{split}
\end{align}
In contrast to beam gain metrics, the beam-matching error is a more stringent constraint \cite{beamMactch}. The actual volumetric of the radar target will considered instead of the original point modeling. Discrete sampling of the continuous beam width requirement, the new metric can be expressed as:
\begin{align}
       \begin{split}
		\varepsilon^{\text{\uppercase\expandafter{\romannumeral2}}}= \sum_{m=1}^{M}{\left| \rho (\theta_m)-  \xi^{\text{I}}(\theta_m)  \right|^2} \label{BME}.
        \end{split}
\end{align}

As observed from Eq.~\eqref{BME}, the goal of beam-matching error is to design beamforming in a way that achieves a beam pattern more similar to a given desired beam pattern $\{\rho (\theta_m)\}_{m=1}^{M}$, where $\theta_m$ represents the angle sample, and $M$ denotes the number of sample points. Typically, the set of angle samples $\{ (\theta_m)\}_{m=1}^{M}$ includes $\theta_p$, which implies $\theta_p \in \{ (\theta_m)\}_{m=1}^{M}, \forall{p}=1,\dots,P$. The number of the sampling points and discrete intervals can be set according to the following guidelines: the larger the object or the less clear the position information, the wider the beam is required for ISAC system. 
	
\section{Green Beamforming Design}
In this section, we jointly design the information beamforming vector and the radar covariance matrix to minimize the transmit power while meeting the radar and communication constraints.
\subsection{Problem Formulation}

Radar constraints can be generated based on different radar metrics. By utilizing the beam-matching error $\varepsilon^{\text{\uppercase\expandafter{\romannumeral2}}}$, the constraints can be written as: $\sum_{m=1}^{M}{\left| \rho (\theta_m)- \xi^{\text{I}}(\theta_m) \right|^2} \le \eta$, where $\eta$ denotes the tolerance error. To address the non-convex constraint, we limit the error for each sampling angle $\theta_m$. Consequently, the constraint can be represented as:
\begin{align}
       \begin{split}
		\rho (\theta_m)-\eta_m \le\xi^{\text{\uppercase\expandafter{\romannumeral1}}}(\theta_m)\le\rho (\theta_m)+\eta_m, \forall m=1,\dots,M,\label{radarCon}
        \end{split}
\end{align}
where $\eta_m$ refers to the tolerance error of angle $\theta_m$. Observing Eq. \eqref{radarCon}, when the $\{\rho (\theta_m)\}_{m=1}^{M}=\{\rho (\theta_p)\}_{p=1}^{P}$, the constrains in Eq. \eqref{radarCon} reduce to constraints on beam gain metrics, which can be expressed as:
\begin{align}
       \begin{split}
		\rho (\theta_p)-\eta_p \le\xi^{\text{\uppercase\expandafter{\romannumeral1}}}(\theta_p)\le\rho (\theta_p)+\eta_p, \forall p=1,\dots,P.\label{radarCon1}
        \end{split}
\end{align}
In the following sections, Eq. \eqref{radarCon1} will be treated as a special case of Eq. \eqref{radarCon}. The corresponding optimization problem can then be formulated as:
	\begin{align}
		(\text{P1}): \mathop {\min }\limits_{{\textbf{\emph{w}}}_k, \textbf{\emph{R}}_d} \quad &\sum_{k=1}^{K}{||{\textbf{\emph{w}}}_k||^2}+\text{tr}(\textbf{\emph{R}}_d) \label{X}\\
		s.t \quad & R_k>R_{\text{min}},\tag{\ref{X}{a}} \forall{k=1,\dots,K}  \label{Xa}\\
		&\rho (\theta_m)-\eta_m \le\xi^{\text{\uppercase\expandafter{\romannumeral1}}}(\theta_m)\le\rho (\theta_m)+\eta_m,\nonumber \\& \forall m=1,\dots,M.\tag{\ref{X}{b}} \label{xb}
	\end{align}
The objective of (P1) is to minimize the total transmit power while satisfying the communication and radar constraints represented by Eq. \eqref{Xa} and Eq. \eqref{xb}, respectively. Additionally, $R_{\text{min}}$ is the  required minimum information rate for all communication users.
\subsection{SDR Approach}
Due to the non-convex nature of the radar constraints in Eq. \eqref{xb}, we employ the techniques of SDR to address this issue. By defining $\textbf{\emph{W}}_k = {\textbf{\emph{w}}}_k{\textbf{\emph{w}}}^{\text{H}}_k \in \mathbb{C}^{N\times N}, \forall k = 1,\dots,K$, $\textbf{\emph{H}}_k = {\textbf{\emph{h}}}_k(\theta_K){\textbf{\emph{h}}}^{\text{H}}_k(\theta_K) \in \mathbb{C}^{N\times N}, \forall k = 1,\dots,K$, and $\textbf{\emph{H}}_m = {\textbf{\emph{h}}}_m(\theta_m){\textbf{\emph{h}}}^{\text{H}}_m(\theta_m) \in \mathbb{C}^{N\times N}, \forall m = 1,\dots,M$, the problem can be equivalently reformulated as:
\begin{align}
    (\text{P2}): \mathop {\min }\limits_{{\textbf{\emph{W}}}_k, \textbf{\emph{R}}_d} \quad &\sum_{k=1}^{K}{\text{tr}(\textbf{\emph{W}}_k)}+\text{tr}(\textbf{\emph{R}}_d) \label{X2}\\
    s.t \quad &    \text{tr}\left(\textbf{\emph{H}}_k\textbf{\emph{W}}_k\right ) -\bar{R} \sum_{i=1,i\ne k}^{K} {\text{tr}\left(\textbf{\emph{H}}_k\textbf{\emph{W}}_i\right )}\nonumber \\&  -\bar{R} \text{tr}(\textbf{\emph{H}}_k \textbf{\emph{R}}_d)\ge \bar{R} \sigma_k^2, \forall k=1,\dots,K,
    \tag{\ref{X2}{a}}  \label{X2a}\\
    &\rho (\theta_m)-\eta_m \le\xi^{\text{\uppercase\expandafter{\romannumeral1}}}(\theta_m)\le\rho (\theta_m)+\eta_m,\nonumber \\& \forall m=1,\dots,M,\tag{\ref{X2}{b}} \label{X2b}\\&
    \textbf{\emph{R}}_d \succeq 0; \textbf{\emph{W}}_k \succeq0 ,\forall k=1,\dots,K \tag{\ref{X2}{c}}\label{X2c} \\&
     \text{Rank}(\textbf{\emph{W}}_k)=1, \forall k=1,\dots,K, \tag{\ref{X2}{d}} \label{Rank_one}
\end{align}
where $\xi^{\text{\uppercase\expandafter{\romannumeral1}}}(\theta_m) = \text{tr}\left(\sum_{k=1}^{K}{\textbf{\emph{H}}_m \textbf{\emph{W}}_k}\right) + \text{tr}\left( \textbf{\emph{H}}_m \textbf{\emph{R}}d \right)$ and $\bar{R} = 2^{R{\text{min}}} - 1$. In (P2), the optimization variables change from $\textbf{\emph{w}}_k$ and $\textbf{\emph{R}}_d$ to $\textbf{\emph{W}}_k$ and $\textbf{\emph{R}}_d$. Unfortunately, the problem (P2) remains non-convex due to the strict rank-one constraints in Eq. \eqref{Rank_one}. By dropping the rank-one constraints of $\textbf{\emph{W}}_k$, we obtain the SDR problem, which can be expressed as:
\begin{align}
    (\text{P3}): \mathop {\min }\limits_{{\textbf{\emph{w}}}_k, \textbf{\emph{R}}_d} \quad &\sum_{k=1}^{K}{\text{tr}(\textbf{\emph{W}}_k)}+\text{tr}(\textbf{\emph{R}}_d) \label{X3}\\
    s.t \quad &\eqref{X2a}, \eqref{X2b}, \text{and}  \eqref{X2c}. \nonumber
\end{align}
\subsection{IRM Algorithm}
\begin{algorithm}[!t]
		\renewcommand{\algorithmicrequire}
         {\textbf{Input:}}
		\renewcommand{\algorithmicensure}{\textbf{Output:}}
		\caption{ISAC Green Beamforming Design}
		\footnotesize
		\begin{algorithmic}[1]
			\REQUIRE ~~\\
                (1) Channel information ${\textbf{\emph{h}}}_k(\theta_k), \forall k=1,\dots,K$;\\
                (2) Tolerance error $\eta_m, \forall m=1,\dots,M$;\\
                (3) Desired beampattern $ \rho (\theta_m),\forall m=1,\dots,M$ ;\\
                (4) Minimum required information rate $R_{\text{min}}$;\\
			    (5) The step size of each iteration $L_{\text{step}}$;	\\
                (6) The threshold of the rank constraint $\varrho$.	
			\ENSURE ~~\\
                (1) Information transmission beamforming vector $\textbf{\emph{w}}_k$;\\
                (2) Radar covariance matrix  $\textbf{\emph{R}}_d$.
			
			\STATE Initialise $\varphi ^{\{0\}}=1$ $j=0$, $r=1$, $L_{\text{step}}=1.5$;
			\STATE Obtain the initial solutions $ \textbf{\emph{R}}^{\{0\}}_d$ and ${\textbf{\emph{W}}}^{{\{0\}}}_k$ by solving (P3);
			\REPEAT
			\STATE Obtain the  ${{\textbf{\emph{V}}}^{\{j\}}_{k}}$ through
 eigenvalue decomposition of  ${\textbf{\emph{W}}}^{{\{j\}}}_k$;
			\STATE $j=j+1$;
			\STATE Obtain $j$-th solutions $ \textbf{\emph{R}}^{\{j\}}_d$ and ${\textbf{\emph{W}}}^{{\{j\}}}_k$ and $r$ by solving (P4);
			\STATE $\varphi ^{\{j\}}=L_{\text{step}}*\varphi ^{\{j-1\}}$;
			\UNTIL  $|r|^2\le \varrho$;
			\STATE Obtain $\textbf{\emph{w}}_k$ through eigenvalue decomposition of ${\textbf{\emph{W}}}^{{\{j\}}}_k$ and  $ \textbf{\emph{R}}_d= \textbf{\emph{R}}^{\{j\}}_d$.
		\end{algorithmic}
	\end{algorithm}
The problem (P3) is now convex and can be solved using numerical tools such as CVX. Upon solving (P3), we obtain the initial solutions, denoted as $\textbf{\emph{R}}^{\{0\}}_d$ and ${\textbf{\emph{W}}}^{{\{0\}}}_k$. If ${\textbf{\emph{W}}}^{{\{0\}}}_k$ precisely satisfies the rank-one constraint, they are already the optimal solutions. Otherwise, additional steps are necessary for forcing the  ${\textbf{\emph{W}}}^{{\{0\}}}_k$ to satisfy the rank-one constraint.  The corresponding $N$ eigenvalues of ${\textbf{\emph{W}}}^{{\{0\}}}_k$  can be expressed as $\zeta_1,~\dots,~\zeta_N$, where  $\zeta_1\leq,\dots,\leq \zeta_{N-1}\leq\zeta_N$ is assumed for convenience. Furthermore, the ${{\textbf{\emph{V}}}^{\{0\}}_{k}}\in \mathbb{R}^{N\times (N-1)}$ is defined to represent the eigenvector matrix for the eigenvalues $\zeta_1$ to $\zeta_{n-1}$, where we can obtain: $\text{diag}(\zeta_1,~\dots,~\zeta_{n-1})= {{\textbf{\emph{V}}}^{\{0\}}_{k}}^{\text{H}}{\textbf{\emph{W}}}^{{\{0\}}}_k{\textbf{\emph{V}}}^{\{0\}}_k$.  In summary, we define a positive relaxation factor $r$ and make the following inequality Eq.~\eqref{IRM_constraint} hold to force the original problem  to have the solution satisfying the rank-one constraint.
\begin{align}
       \begin{split}
		r {\textbf{\emph{I}}}_{N-1}- {{\textbf{\emph{V}}}^{\{0\}}_{k}}^{\text{H}}{\textbf{\emph{W}}}^{{\{0\}}}_k{\textbf{\emph{V}}}^{\{0\}}_k\succeq 0, \label{IRM_constraint}
        \end{split}
\end{align}
where ${\textbf{\emph{I}}}_{N-1}$ is a unit matrix of order $N-1$. As the positive number $r$ approaches zero, the $N-1$ smallest eigenvalues are forced to be zero, implying that ${\textbf{\emph{W}}}^{{\{0\}}}_k$ satisfies the rank-one constraints. The key idea of the IRM is to progressively reduce the value of $r$ through iterations. We assume the $j$-th step iteration of the IRM problem is formulated as:
\begin{align}
    (\text{P4}): \mathop {\min }\limits_{{\textbf{\emph{W}}}^{\{j\}}_k, \textbf{\emph{R}}^{\{j\}}_d}
    \quad &\sum_{k=1}^{K}{\text{tr}(\textbf{\emph{W}}^{\{j\}}_k)}+\text{tr}(\textbf{\emph{R}}^{\{j\}}_d) +\varphi ^{\{j\}}r \label{X4}\\
    s.t \quad &    \text{tr}\left(\textbf{\emph{H}}_k\textbf{\emph{W}}^{\{j\}}_k\right ) -\bar{R} \sum_{i=1,i\ne k}^{K} {\text{tr}\left(\textbf{\emph{H}}_k\textbf{\emph{W}}^{\{j\}}_i\right )}\nonumber \\&  -\bar{R} \text{tr}(\textbf{\emph{H}}_k \textbf{\emph{R}}^{\{j\}}_d)\ge \bar{R} \sigma_k^2, \forall k=1,\dots,K,
    \tag{\ref{X4}{a}}  \label{X4a}\\
    &\rho (\theta_m)-\eta_m \le\xi^{\text{\uppercase\expandafter{\romannumeral1}},{\{j\}}}(\theta_m)\le\rho (\theta_m)+\eta_m,\nonumber \\& \forall m=1,\dots,M,\tag{\ref{X4}{b}} \label{X4b}\\&
    \textbf{\emph{R}}^{\{j\}}_d \succeq 0; \textbf{\emph{W}}^{\{j\}}_k \succeq0 ,\forall k=1,\dots,K \tag{\ref{X4}{c}}\label{X4c}\\&
    r {\textbf{\emph{I}}}_{n-1}- {{\textbf{\emph{V}}}^{\{j-1\}}_{k}}^{\text{H}}{\textbf{\emph{W}}}^{{\{j\}}}_k{\textbf{\emph{V}}}^{\{j-1\}}_k\succeq 0.\tag{\ref{X4}{d}}\label{X4d}
\end{align}
where $\varphi ^{\{j\}}$ is the weight of $j$-th iteration, and $\xi^{\text{\uppercase\expandafter{\romannumeral1}},\{j\}}(\theta_m)=\text{tr}\left(\sum_{k=1}^{K}{\textbf{\emph{H}}_m \textbf{\emph{W}}^{\{j\}}_k}\right)+\text{tr}\left( \textbf{\emph{H}}_m \textbf{\emph{R}}^{\{j\}}_d \right)$. In each iteration, we need to keep increasing the value of  $\varphi ^{\{j\}}$ to increase the weight of the $r$ in the objective function. In that case, the objective function wants to minimise the value of $r$, which will cause ${\textbf{\emph{W}}}^{{\{j\}}}_k$ to guradually satisfy the rank-one constraint.  The effectiveness and convergence of this approach has been demonstrated in \cite{IRM_iter}.
For clarity, the IRM iteration algorithm is summarised in Algorithm 1.

Using the interior point method \cite{complexity}, (P4) can be solved with a complexity of $\mathcal{O}\left( N^7\right)$. Assuming that $\delta$ iterations are required for Algorithm 1 to converge, the total complexity becomes $\mathcal{O}\left(\delta N^7\right)$.

	\section{Numerical Results}
	In this section, we present the numerical results for our proposed beamforming design focused on power minimization. The simulation outcomes validate  the effectiveness of the proposed algorithm for beamforming design in ISAC systems. The simulation parameters are summarized in TABLE \ref{table1}.

\begin{table}[H]
		\centering
		\renewcommand{\arraystretch}{1.3} 
		\caption{Simulation Parameters}
		\label{table1}
	\begin{tabular}{c|c}
		\hline
		Parameter                & Value                 \\ \hline\hline
		Carrier frequency $f_{\rm{c}}$ (GHz)   & 0.95          \\ \hline
		Transmit and receive gain $G_{\text{T}}$ and $G_{\text{R}} ~$(dB)  & 0   \\\hline
	    Distance of communication user $d_{k} ~($\text{m}$)$         & [10-20] \\ \hline
        Distance of radar target $d_{p} ~($\text{m}$)$               & [10-20] \\ \hline
		Number of antennas $N$          & [10-20]                    \\ \hline
        Power of noise $\theta_k~\forall k=1,\dots.K$ (dBm)         & -75                    \\ \hline
	\end{tabular}
    \end{table}
	
    The desired beampattern with beam width $\Delta$ is defined as :

    \begin{align}
		\begin{split}
		  \rho(\theta)= \phi_p,  \quad\theta_p-\frac{\Delta}{2}\leq\theta\leq\theta_p+\frac{\Delta}{2},\quad \forall p=1,\dots,P,
		\end{split}
    \end{align}
    where $\phi_p$ is related to the distance of radar target $d_p$\cite{Beam_match}, which can be further expressed as $\phi_p=\Gamma-10\log_{10}{\frac{G_{\text{T}}G_{\text{R}}{\lambda}^2}{(4\pi)^2 d_p^2}}$, with the minimum power of the ISAC signal at the radar target typically  assumed to be $\Gamma =-13~\text{dBm}$\cite{UAV_beamGain}.

\begin{figure}[!t]
		\centering
		\includegraphics[width=0.45\textwidth]{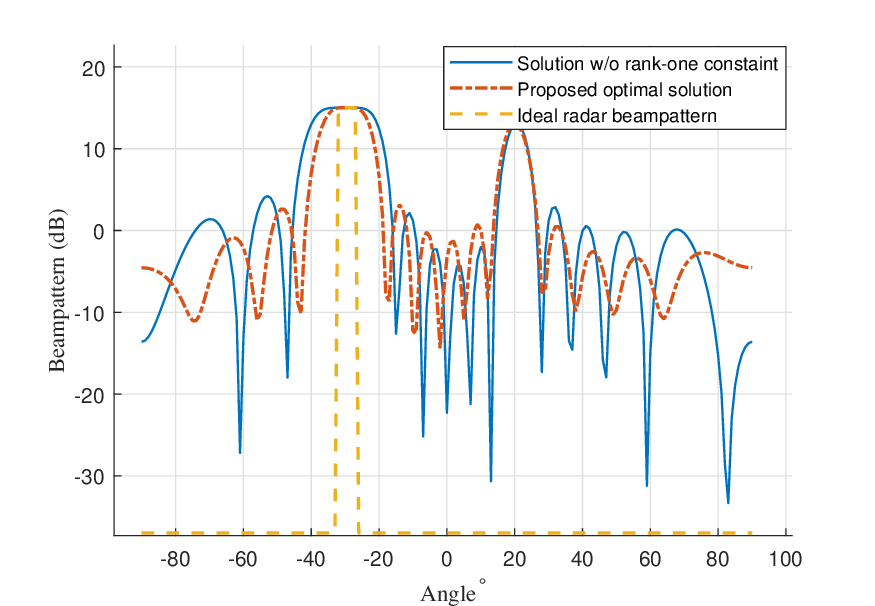}
		\caption{Beampattern under different solution with $N=15$ antennas and a beam width of $\Delta=5^\circ$. The radar target is positioned at an angle of $-30^\circ$ at a distance of 20m, and the communication user is located at $20^\circ$ with a distance of 20m.} \label{IRM}
	\end{figure}
 
    Fig. \ref{IRM} illustrates the beampattern difference between the proposed optimal solution and the solution obtained without the rank-one constraint, which is derived by solving the SDR problem (P3). While the without-the-rank-one solution has more degrees of freedom due to the absence of the rank-one constraint in (P3), this solution serves as a lower bound on the power consumption of the ISAC beamforming design. The proposed optimal solution, obtained using Algorithm 1, consumes a total power of $8.15~\text{dB}$, which is identical to that of the without-the-rank-one solution, thus confirming the efficacy of our proposed algorithm.

      \begin{figure}[!t]
		\centering
		\includegraphics[width=0.45\textwidth]{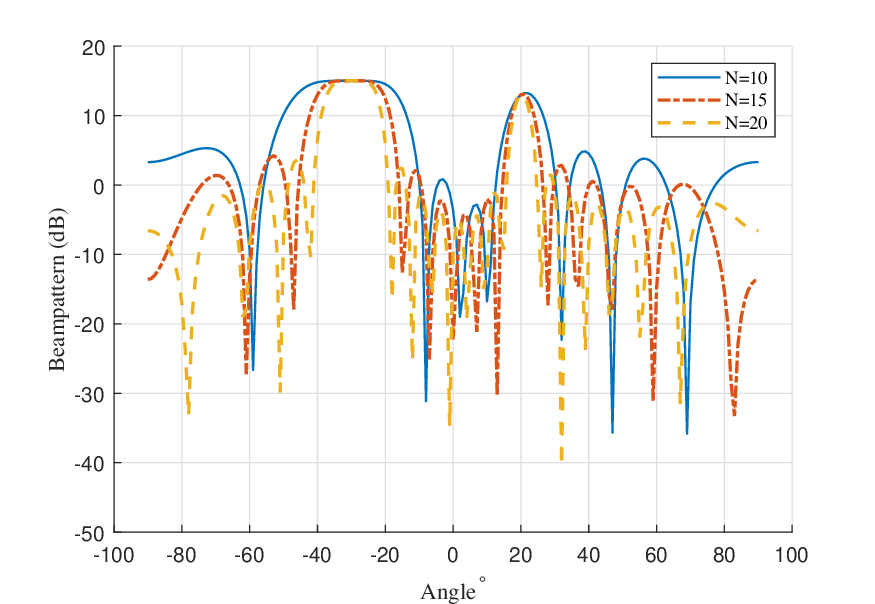}
		\caption{Beampattern under the different number of antennas with $\Delta=5^\circ$.  The radar target is positioned at an angle of $-30^\circ$ at a distance of 20m, and the communication user is located at $20^\circ$ with a distance of 20m.} \label{antenna}

	\end{figure}

    \begin{figure}[H]
\vspace{-6pt}
\subfloat[]{\includegraphics[width=0.45\textwidth]{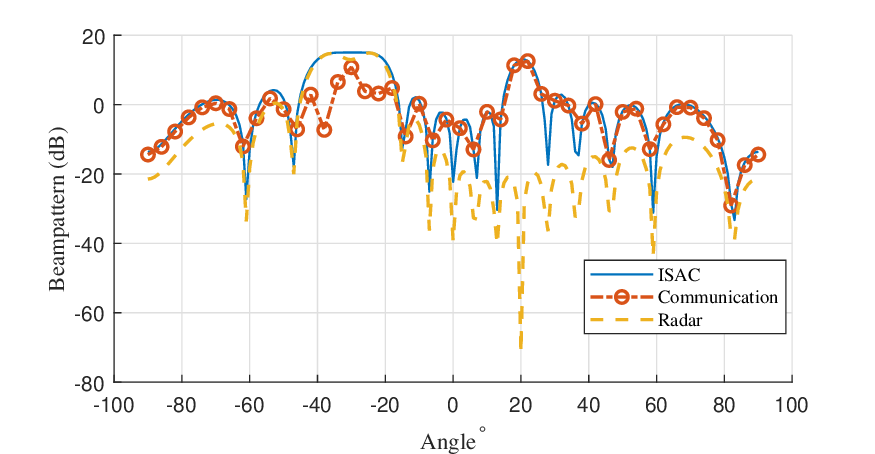}%
\label{sub_figure1}}\\
\vspace{-3pt}
\subfloat[]{\includegraphics[width=0.45\textwidth]{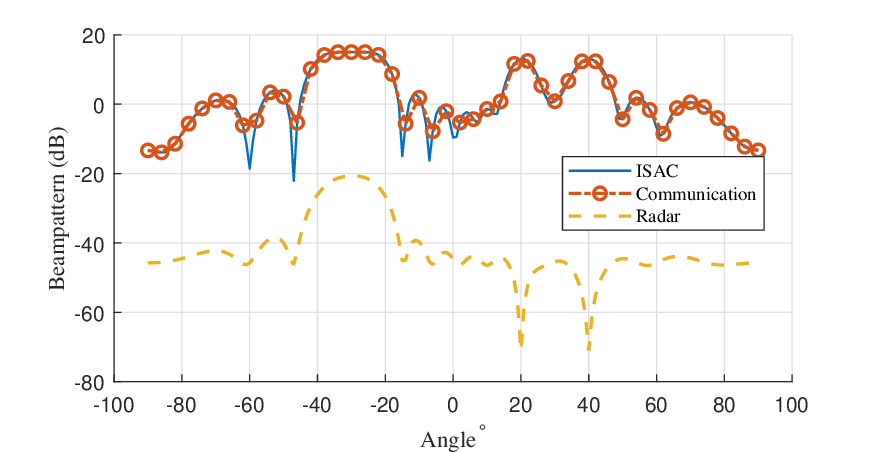}%
\label{sub_figure2}}\\
\vspace{-3pt}
\subfloat[]{\includegraphics[width=0.45\textwidth]{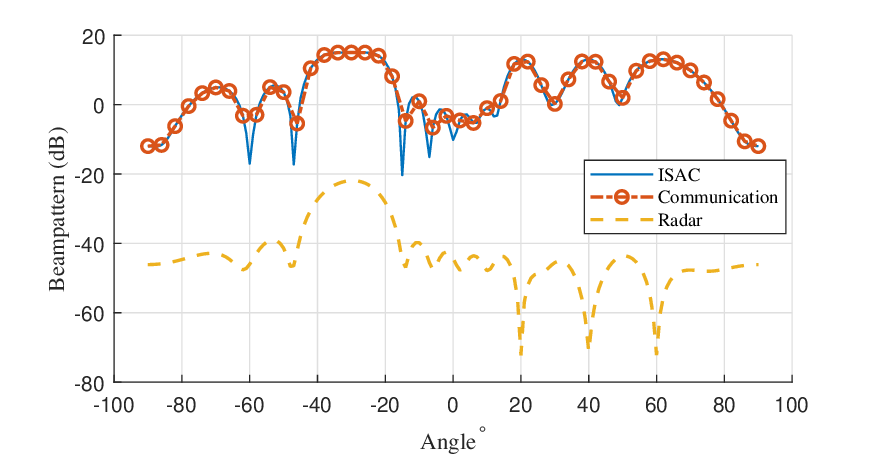}%
	\label{sub_figure3}}\\
 \vspace{-3pt}
\subfloat[]{\includegraphics[width=0.45\textwidth]{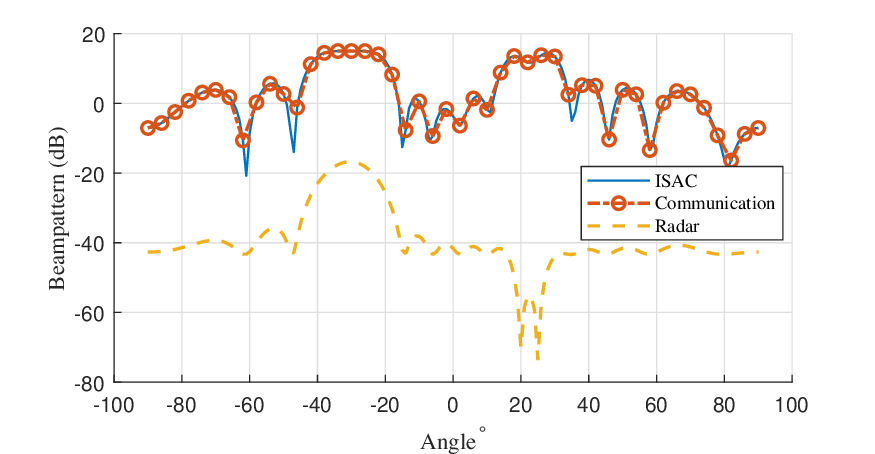}%
	\label{sub_figure4}}\\
 \vspace{-3pt}
\subfloat[]{\includegraphics[width=0.45\textwidth]{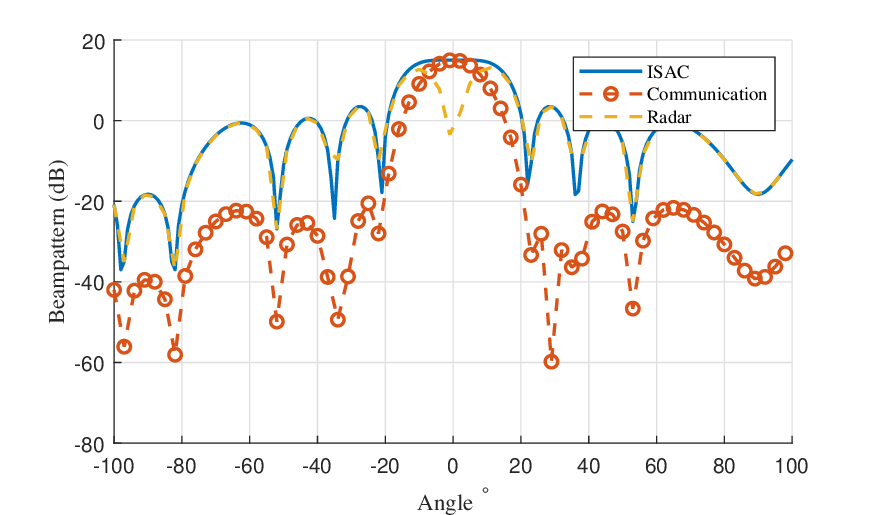}%
\label{sub_figure5}}
\vspace{-3pt}
\caption{The beampattern for the different angle of the communication user, where $N=10$ and $\Delta=5^\circ$.}
\label{UU}

\end{figure}

In Fig. \ref{antenna}, we investigate the impact of the number of antennas on the beampattern. As the number of antennas increases, the beams become narrower, resulting in lower power consumption in the side lobes. In comparison to the system with $N=20$, the system with $N=10$ has wider side lobes, which means it consumes more useless power in the side lobes. Particularly, in the radar area beampattern, more antennas can provide more degrees of freedom to minimize the power of side lobes. The total power of the system with $N=10$ is $9.8~\text{dB}$, while the system with $N=20$ has a total power of $6.9~\text{dB}$, demonstrating the effectiveness of increasing the number of antennas.

Fig. \ref{UU} depicts the correlation between the communication and radar waveform components for various numbers and angles of users. The "ISAC" line in the graph shows the integrated waveform, whereas the "Communication" and "Radar" lines represent the communication waveform component and radar waveform component, respectively. The five subgraphs are set up as follows:
 (a) One communication user is located at $20^\circ$, while one radar target at $-30^\circ$. (b) Two communication users are located at $20^\circ$ and $40^\circ$, while one radar target at $-30^\circ$.  (c) Three communication users are located at $20^\circ$, $40^\circ$ and $60^\circ$, while one radar target at $-30^\circ$. (d) Two communication users are located at $20^\circ$ and $25^\circ$, while one radar target at $-30^\circ$. (e) One communication user is located at $0^\circ$, while one radar target at $0^\circ$.
The performance of the radar waveform component is better for the radar task in Fig. \ref{UU}(a) than in Fig. \ref{UU}(b) due to a single communication user's inability to meet the beam width of the radar target when there is a rank-one constraint. Additionally, the radar component is almost zero at the communication user's angle, avoiding interference with communication. When comparing Fig. \ref{UU}(b) and Fig. \ref{UU}(d), the closer the angles of communication users 1 and 2, the higher the internal interference between them, requiring more power to eliminate it. The total power consumption in the setting of Fig. \ref{UU}(b) is $10.5~\text{dB}$, while that in Fig. \ref{UU}(d) is $12~\text{dB}$. In Fig.~\ref{UU}(e), in order to avoid the interference  with communications, the radar component at $0^\circ$ as low as possible. More, the sensing task is shared by the communication component and the radar component.

\begin{figure}[H]
		\centering
		\includegraphics[width=0.45\textwidth]{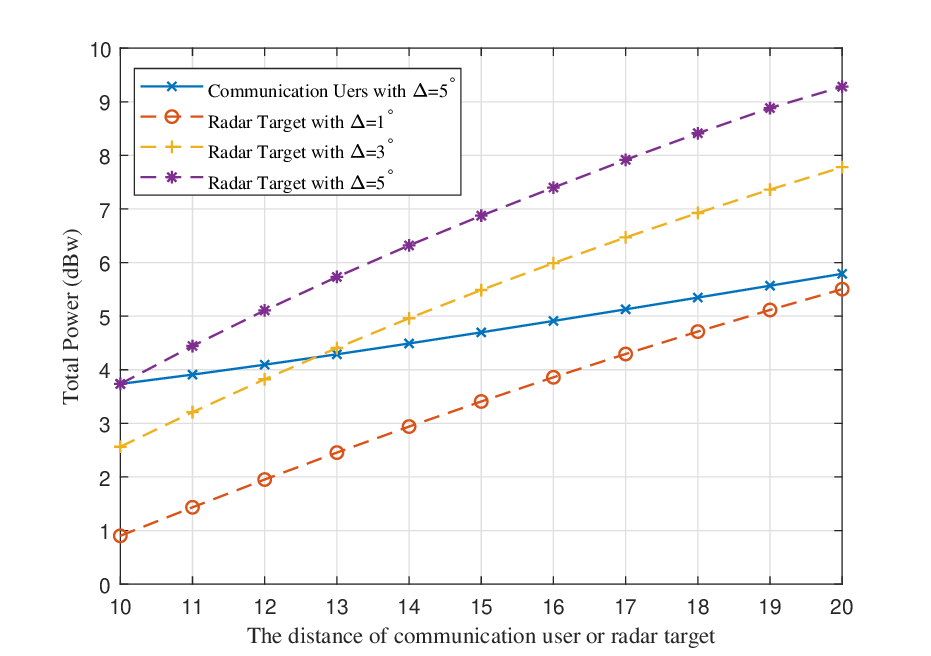}
		\caption{The total power consumption versus the distance of communication user and radar target with different beam width $\Delta$, where the number of antennas $N=10$} \label{ditance}
 
	\end{figure}

The relationship between the distance and the total power consumption is depicted int Fig.~\ref{ditance}, where the effect of the beam width is focused on. In this scenario, there is a single communication user is placed at $20^\circ$ while the signal radar target is set at $-30^\circ$. Employing the control variable method, the distance of the communication user is fixed to  $10~\text{m}$ when the radar target are moved form $10~\text{m}$ to $20~\text{m}$ and vice versa. The results demonstrate that an increase in the distance to either the communication user or the radar target result in an increase in total power consumption. Furthermore, it is noted that the radar target's distance has a greater impact on the total power consumption than that of the communication user, which is because the beam width constraint of the radar requires more power in the ISAC system. Comparing the different beam width settings, when the beam width expands from $\Delta=1^\circ$ to $\Delta=5^\circ$, the total power consumption increases nonlinearly.

\vspace{-6pt}

\section{Conclusion}
In conclusion, this paper presents a green beamforming design for ISAC systems, employing beam-matching error as a metric for evaluating radar performance. To address the non-convex challenges in beamforming design, we utilize SDR for tackling the rank-one relaxation problem and subsequently implement the IRM for rank-one recovery. The simulation outcomes highlight the effectiveness of our proposed green beamforming design and underscore the remarkable performance of the radar component in sensing tasks.
\vspace{-6pt}
\bibliography{refer}
\end{document}